\documentclass[a4paper]{extarticle}
\usepackage[utf8]{inputenc}
\usepackage[T1]{fontenc}
\usepackage[backend=biber,backref=true]{biblatex}
\usepackage{parskip} 
\usepackage{graphicx}
\usepackage{fullpage}
\usepackage{color}
\usepackage{amsmath}
\usepackage{amsthm}
\usepackage{amssymb}
\usepackage{siunitx}
\usepackage{multicol}
\usepackage{blkarray}
\usepackage{comment}
\usepackage[section]{placeins}
\usepackage{listings}
\usepackage{arrayjobx}
\usepackage{array}
\usepackage{enumitem}
\usepackage{mathtools}
\usepackage{setspace}
\usepackage{algorithm}
\usepackage{algpseudocode}
\usepackage{tikz}
\usetikzlibrary{calc,shapes,arrows.meta,decorations.markings}
\newcommand{\tikzmark}[1]{\tikz[overlay,remember picture] \node (#1) {};}

\DeclareRobustCommand\CXX{C\kern-.05em {\scalebox{0.9}{\textbf{+\kern-.10em+}}}}
\DeclareRobustCommand\CC{C\texttt{++}}
\def\bftab{\fontseries{b}\selectfont}
\newtheorem*{theorem}{Theorem}

\begin{document}
\title{The iisignature library: efficient calculation of iterated-integral signatures and log signatures}
\author{Jeremy Reizenstein\\Alan Turing Institute and Centre for Complexity Science, University of Warwick\thanks{Supported by the Engineering and Physical Sciences Research Council and The Alan Turing Institute under the EPSRC grant EP/N510129/1}
\\Benjamin Graham, Facebook AI Research %
}
\date{February 2018}
\maketitle

\def\ii{{\texttt{iisignature}}}
\def\pypi{{\texttt{PyPI}}}
\def\numpy{{\texttt{numpy}}}
\def\scipy{{\texttt{scipy}}}
\def\i#1{\index{#1@\texttt{#1}}}
\def \hilite#1{\underline{\color{blue}\textbf{#1}}}
\def \alph#1{{\color{blue}\mathbf{#1}}}
\def \lex{<_L}
\begin{abstract}
Iterated-integral signatures and log signatures are
vectors calculated from a path that characterise its shape. They come from the theory of differential equations driven by rough paths, and also have applications in statistics and machine learning. 
We present algorithms for efficiently calculating these signatures, and benchmark their performance. We release the methods as a Python package.
\end{abstract}
\section{Introduction}
\ii\ is a Python package which calculates the iterated-integral signatures and log signatures of paths. 
The signature is an object which is crucial in the mathematical theory of rough paths, and the calculations have proved to be useful in machine-learning applications, particularly classification problems where the data itself is a stream or a path in space, ranging from an application to online Chinese handwriting recognition in 2013 \cite{BEN} to skeleton-based human action recognition in 2017 \cite{action}. Other domains where the data has this form include signals from EEG and other medical monitors, sound and financial time series, where some set of numbers is varying in time. Often the samples can be noisy, can have varying length and both local and global structure can be important. A survey of such applications is given in \cite{OxSigIntro}. 

An existing open-source implementation is the \verb|esig| package from CoRoPa\cite{coropa}.
CoRoPa operates in a sparse fashion, keeping track of only non-zero elements of the signature. 
This has been known as 
\emph{sparse signatures}.
It is useful in some applications of signatures in high-dimensional spaces where the path only moves in certain combinations of the input dimensions.

The particular focus of \ii\ is piecewise-linear fixed-dimensional paths which typically move in all their dimensions. 
In this setting, usually none of the elements of the signature are zero. Sparse methods impose a significant overhead in this context; \ii\ is directed at these \emph{dense} signatures. 

We study the mathematical properties of the free Lie algebra to implement algorithms for calculating signatures in the dense case. We also benchmark the performance of these algorithms, and provide an efficient open-source implementation. 
This paper is organised as follows. 
The rest of this section introduces signatures, log signatures and the library. 
Signature algorithms are discussed in \ref{sec:sigs}. Log signature methods are introduced in \ref{sec:logsigs}, the direct method is discussed in \ref{sec:c} and the projection method in \ref{sec:s}. 
Considerations around the implementation are presented in \ref{sec:impl}. 
Indicative timings are given in \ref{sec:time} and memory usages in \ref{sec:mem}. 
We briefly discuss other functionality provided by the library in \ref{sec:other} before concluding.

\ii\ is hosted at \url{https://github.com/bottler/iisignature} and is available on \pypi. 

\subsection{What is the signature of a path?}

The iterated-integral signature of a continuous path is an infinite sequence of numbers. It is used in the mathematical theory of differential equations driven by paths. In these problems, a path is the driving signal for a certain type of system. It turns out that the signature is exactly the information about a path which you need to know in order to predict how the output of the system will behave, using a generalisation of Taylor's theorem. 
It is natural that the signature would also be the right information to extract from a path if we want a machine-learning algorithm to understand the shape of the path.

In general, a $d$-dimensional continuous path is given by a function 
from an interval $[a,b]\subset\mathbb{R}$ to $\mathbb{R}^d$. 
Its signature depends on the appearance of the path and the direction it was created, but not the speed at which it was created. If a path is modified by adding or removing a section which is exactly backtracked over, then its signature does not change. If the path has a time dimension along which it always increases (for example it is the graph of a function of time) then exact backtracking is impossible and so any two different paths will have different signatures.%
\footnote{For paths which are continuously differentiable at all but finitely many places, such as the paths which we deal with in \ii\ which are piecewise linear between a set of specified points, the signatures of two different paths which contain no exact backtracking will differ.\cite{chen}}

The signature is divided into units called levels. We cannot store the whole signature of a path on a computer, rather we calculate a certain number of levels of it. 
The more levels of a signature are known, the more precisely the shape of the path is determined.
If a path changes very slightly, the first few levels of its signature will also only change very slightly. If a path is moved (translated) but retains its shape, its signature will not change. 

The number of elements of level $m$ of the signature of a $d$-dimensional path is $d^m$. They are the values of iterated integrals which consist of $m$ nested integrals, and they are labelled with $m$ numbers each corresponding to one of the dimensions. To distinguish these numbers which label the dimensions from other numbers, we write them bold and in blue. For example, a two-dimensional path might be given in coordinates as  $(\gamma_{\alph1}(t),\gamma_{\alph2}(t))$ as $t$ varies from $a$ to $b$. Its signature is a function denoted by $X^\gamma_{a,b}$. Level three of its signature has eight elements, called $X_{a,b}^\gamma(\alph{111})$, $X_{a,b}^\gamma(\alph{112})$ and so on. The one indexed by the word $\alph{122}$ is 
\begin{align}
X_{a,b}^\gamma(\alph{122})=\int_{t_1=a}^b\int_{t_2=a}^{t_1}\int_{t_3=a}^{t_2}d\gamma_{\alph1}(t_3)\,d\gamma_{\alph2}(t_2)\,d\gamma_{\alph2}(t_1).
\end{align}

In general, the signature can be defined inductively on the length of the word. The signature of the empty word is the single value in level 0, and it is defined to always be 1. If $w$ is a word and $i\in\{\alph{1},\alph{2},\dots,\alph{d}\}$ then $X^\gamma_{a,b}(wi)$ is defined as $\int_a^tX^\gamma_{a,t}(w)\,\gamma_i'(t)\,dt$.%
\footnote{Level $m$ can be thought of as taking values in $(\mathbb{R}^d)^{\otimes m}$, which is a $d^m$-dimensional real vector space. In this form, the signature is seen to be an element of the tensor algebra $T(\mathbb{R}^d)=\bigoplus_{m=0}^\infty (\mathbb{R}^d)^{\otimes m}$, the infinite direct sum of tensor powers of $\mathbb{R}$.}

The information in the first level of the signature is the total displacement of the path, i.e.~the direction and distance from its starting point to its ending point. The information which the second level of the signature adds is the \emph{signed area} of the path projected in each plane. Higher levels of the signature provide more detailed information about the path's shape.

\subsection{Signed area}
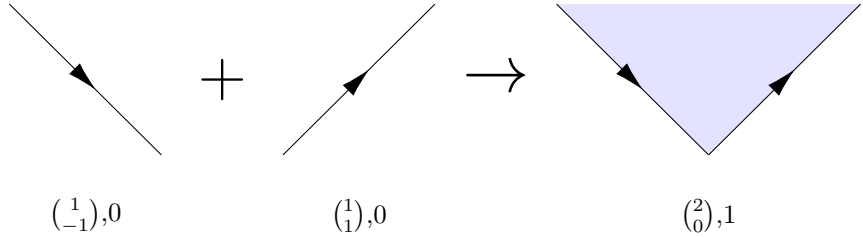
\begin{figure}[H]
	\begin{center}
\begin{tikzpicture}

\begin{scope}[decoration={ 
	markings,
	mark=at position 0.5 with {\arrow[xshift=2mm]{Latex[length=4mm,width=2mm]};}}
] 
\def\halfsize{1}
\def\halfskip{0.8}
\def\bottomrow{-\halfsize-\halfskip}
\def\size{\halfsize+\halfsize}
\def\startTwo{\size+\halfskip+\halfskip}
\def\startThree{\startTwo+\size+\halfskip+\halfskip}
\draw[postaction={decorate}] (0,\halfsize)--(\size,-\halfsize);
\node at (\halfsize,\bottomrow){$\binom{1}{-1}$,0};
\node at (\size+\halfskip,0) {\Huge$+$};
\draw[postaction={decorate}] (\startTwo,-\halfsize)--(\startTwo+\size,\halfsize);
\node at (\startTwo+\halfsize,\bottomrow){$\binom{1}{1}$,0};
\node at (\startTwo+\size+\halfskip,0) {\Huge$\to$};
\fill [color=blue!11]
(\startThree,\halfsize)--(\startThree+\size,-\halfsize)--(\startThree+\size+\size,\halfsize)--cycle;
\draw[postaction={decorate}] (\startThree,\halfsize)--(\startThree+\size,-\halfsize);
\draw[postaction={decorate}] (\startThree+\size,-\halfsize)--(\startThree+\size+\size,\halfsize);
\node at (\startThree+\size,\bottomrow){$\binom{2}{0}$,$1$};
\end{scope}
\end{tikzpicture}
		\caption{Concatenating paths and the corresponding total displacements  and total signed areas.\label{fig:sig-comp}}
	\end{center}
\end{figure}

For a two-dimensional path, the information carried by the first two levels of the signature is the total displacement of the path (in the first level, which is two numbers) and the signed area between the path and the straight line from its beginning to end. 
Figure~\ref{fig:sig-comp} shows this information for two straight lines and their combination, which contains area.

The following is an intuitive definition of the signed area of a path in the plane. For a closed path, that is one which ends where it starts, the signed area is the sum of the signed areas of the regions bounded by the path, which is the area times the number of times the path goes round that region in an anticlockwise manner minus the number of times the path goes round it clockwise (i.e.~the winding number). For example, in the path shown in Figure~\ref{fig:winding}(a), regions whose areas count positively are labelled with a \raisebox{1mm}{\tiny\boldmath\color{red}$+$}, and negatively with a \raisebox{1mm}{\tiny\boldmath\color{red}$-$}. One region's area counts twice negatively; it is labelled with \raisebox{1mm}{\tiny\boldmath\color{red}$--$}. For a more general path, its signed area is the signed area of the closed path you get by joining it with a straight line from its end to its start.
\def\windingpoints{(0,0) (1,1) (3,1) (1.5,-1) (1,-1) (0.8,0.7) (1.3,1.6)(1.1,1.7)(0.9,1.4)(1.8,0.5)(2.8,-1.3)(3.6,0.2)}
\begin{figure}[H]
	\begin{center}
		\begin{tikzpicture}[font=\tiny\boldmath\color{red}]
		\begin{scope}[decoration={
			markings,
			mark=at position 0.08 with {\arrow{Latex[length=2mm]};},
			mark=at position 0.3 with {\arrow{Latex[length=2mm]};},
			mark=at position 0.52 with {\arrow{Latex[length=2mm]};},
			mark=at position 0.68 with {\arrow{Latex[length=2mm]};},
			mark=at position 0.9 with {\arrow{Latex[length=2mm]};} 
		}
		] 
		\node at (-0.4,1) {\normalsize\color{black}(a)};
		\draw [->,postaction={decorate}] plot [smooth cycle] coordinates {\windingpoints};
		\node at (0.5,.3) {$-$};
		\node at (1.4,-0.4) {$-$};
		\node at (1.3,0.4) {$--$};
		\node at (2.2,0.7) {$-$};
		\node at (1.07,1.1) {$-$};
		\node at (2.9,-0.4) {$+$};
		\node at (1.05,1.5) {$+$};
		\end{scope}
		\begin{scope}[scale=1, shift={(5,0)}] 
		\begin{scope}[decoration={
			markings,
			mark=at position 0.4 with {\arrow{Latex[length=2mm]};}}
		] 
		\node at (-0.3,1) {\normalsize\color{black}(b)};
		\draw [smooth,tension=1,postaction={decorate}] plot coordinates {(1.1,1.1)(0.5,0.7)(0,0)(1,-1) (1.8,0) (0.7,1.1)};
		\draw [blue, densely dotted] (1.1,1.1) -- (0.7,1.1);
		\node at (0.9,1.05) {$-$};
		\node at (0.9,0) {$+$};
		
		\begin{scope}[shift={(2,0)}]
		\draw [smooth,tension=1,postaction={decorate}] plot coordinates {(1,0.9)(0.5,0.8)(0,0)(1,-1) (2,0) (1.2,1)};
		\draw [blue, densely dotted] (1.2,1) -- (1,0.9);
		\node at (1,0) {$+$};
		\end{scope}
		\end{scope}
		\begin{scope}[shift={(5,0)},decoration={
			markings,
			mark=at position 0.24 with {\arrow{Latex[length=2mm]};}}]
		\draw [smooth, tension=1,postaction={decorate}] plot coordinates{(0.4,0.6)(0.4,0.3)(1,0)(1.5,-0.6)(1,-1)(0.5,-0.6)(1,0)(1.5,0.5)(1,1)(0.5,0.7)};
		\draw [blue, densely dotted] (0.5,0.7) -- (0.4,0.6);
		\node at (0.9,0.5) {$+$};
		\node at (1,-0.5) {$-$};
		\node at (0,1) {\normalsize\color{black}(c)};
		
		\end{scope}
		\end{scope}
		\end{tikzpicture}	
		\caption{\label{fig:winding} (a) A complicated closed path showing the multiplicity of each region it contains, (b) two idealised handwritten digit 0s showing the completion into a closed curve and showing how the nature of the straight line completion has only a small effect on the area, and (c) an illustration of an idealised handwritten digit 8 showing why, although it is a large object, its area might be small due to cancellation of a positive and negative part.}
	\end{center}
\end{figure}
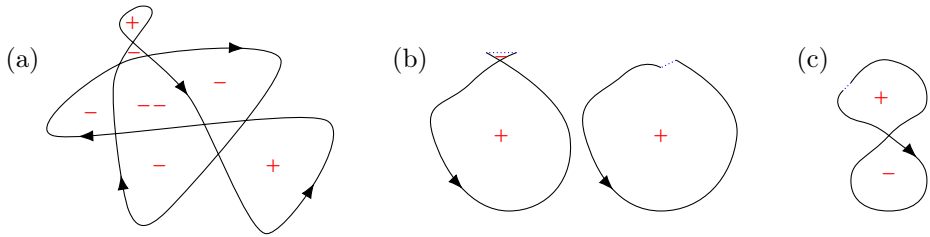

\begin{figure}[H]
	\begin{center}
		\includegraphics[width=4.387in]{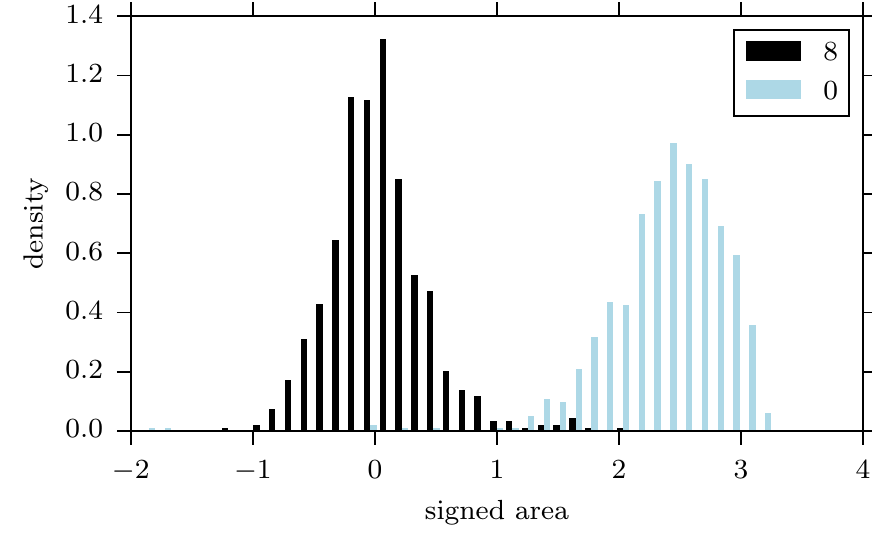}
		\caption{\label{fig:histogram80}Histogram of areas of the first stroke of each 0 and 8 in the training portion of the Pendigits dataset.}
	\end{center}
\end{figure}

As an example of how the area can be useful in classifying the shape of the path, consider classifying handwritten digits 0 and 8. Usually these are written with a single stroke which ends near its beginning, so the displacement is insufficient for distinguishing them. However, the encompassed areas are statistically different. The figure 0 is typically formed from a single anticlockwise loop, generating a positive signed area, while the figure 8 contains two regions with opposite sign, leading to cancellation of signed area. The diagrams in Figure~\ref{fig:winding}(b) and (c) illustrate this. The Pendigits dataset \cite{pendigits} collected the traces of many people writing the digits 0 to 9, and the histogram in Figure~\ref{fig:histogram80} shows how different the  signed areas of the first (and usually only) strokes of these digits are. This clear separation is an illustration of the potential usefulness of the signature for classification.

\subsection{What is the log signature of a path?}
The log signature is a compressed version of the signature. 
It carries the same information, but in a more compact way. 
It is also divided into levels. 
Up to level $m$, the log signature contains fewer numbers than the signature.
Any given set of values for these numbers actually gives the log signature of some path, whereas 
this is not the case for signatures, because there is some redundancy in the signature.
For example the first two levels of the signature of a two-dimensional path consists of $2+2^2=6$ numbers but we saw that this information is the path's total displacement and signed area, which can be stored in three numbers, which are exactly the first two levels of the log signature. In applications, the log signature might be less susceptible to roundoff error. The log signature is defined in terms of the signature, in a way analogous to logarithms of numbers, but can be calculated via an independent algorithm.

\subsection{Using the library}

The library is designed to make calculating large numbers of signatures and log signatures 
fast. 
To this end, preparatory calculations for the log signature calculation happen in a separate preparation function called \verb|prepare|. This also means the library's size on disc can be small; there is no separate code for specific numbers of dimensions and levels.

Figure~\ref{fig:code} shows an example of calculating the signature and log signature of a 3-dimensional path up to level 4, which is specified as a set of $n$ points.

\begin{figure}[H]
\newsavebox{\Lst}
\begin{lrbox}{\Lst}
\begin{lstlisting}[language=Python,keywordstyle=\color{blue},commentstyle=\tt\color{red}]
import iisignature

path = ... #some numpy array of shape (n,3)
signature = iisignature.sig(path,4)
s = iisignature.prepare(3,4)
logsignature = iisignature.logsig(path,s)
\end{lstlisting}
\end{lrbox}
\begin{center}
\fbox{\usebox{\Lst}}
\end{center}
\caption{\label{fig:code}Simple use of \ii.}
\end{figure}

After running this code, \verb|signature| will be a \numpy\ array of the values of levels 1, 2, 3 and 4 of the signature, which has length $3+3^2+3^3+3^4=120$. Note that level 0, which is the constant 1 and contains no information about the path, is excluded from the output of \ii. \verb|logsignature| will be a \numpy\ array of the values of levels 1, 2, 3 and 4 of the logsignature, which has length $3+3+8+18=32$.

\section{Signatures}\label{sec:sigs}

Calculating the signature of a path can be done inductively relying on the following two rules.
\begin{itemize}
	\item If $\gamma$ is a straight line defined on the interval $[a,b]$ then its signature as a function on words is \begin{equation}X^\gamma_{a,b}(i_1i_2\ldots i_m)=\frac1{m!}\prod_{j=1}^m(\gamma_{i_j}(b)-\gamma_{i_j}(a))\label{eq:straightsig}.\end{equation} 
	
	Grouped by levels, using $x=\gamma(b)-\gamma(a)$ as the displacement, the signature looks like
	\begin{equation}
	\left(1,x,\frac{x\otimes x}{2!},\frac{x\otimes x\otimes x}{3!},\dots\right)
	\end{equation}
	where $\otimes$ is the tensor product. Alternatively, if each level is thought of as a vector of numbers, this formula should be read with $\otimes$ denoting the Kronecker product.
	\item If $a<b<c$ then the result (from \cite{chen}) known as \textbf{Chen's identity} states that \begin{equation}
	X^\gamma_{a,c}(i_1i_2\ldots i_m) 
	=\sum_{j=0}^mX^\gamma_{a,b}(i_1i_2\ldots i_{j-1})X^\gamma_{b,c}(i_ji_{j+1}\ldots i_m).\label{eq:chen}
	\end{equation}
	
	Grouped by levels, this signature looks like
	\begin{align}
	\Big(1,X^{(1)}_{a,c},X^{(2)}_{a,c},\dots\Big)=\Big(1,X^{(1)}_{a,b}+X^{(1)}_{b,c},X^{(2)}_{a,b}+X^{(1)}_{a,b}\otimes X^{(1)}_{b,c}+X^{(2)}_{b,c},\qquad\qquad\\ X^{(3)}_{a,b}+X^{(2)}_{a,b}\otimes X^{(1)}_{b,c}+X^{(1)}_{a,b}\otimes X^{(2)}_{b,c}+X^{(3)}_{b,c},\dots\Big)\nonumber
	\end{align}
	
\end{itemize}

When calculating the signature of a path given as a series of straight-line displacements, we start with the signature of the first displacement (calculated from (\ref{eq:straightsig})) and step-by-step concatenate on the signature of each succeeding displacement using (\ref{eq:chen}).

Level $m$ of the signature contains $d^m$ values. Calculating it for a displacement using (\ref{eq:straightsig}) takes $d+d^{m}$ multiplications beyond what has already been calculated for lower levels. However, in the signature of a straight line, each level is a symmetric tensor and so level $m$ only contains $\binom{d+m-1}{m}$ distinct values, using the formula for unordered sampling with replacement. An alternative, more complicated, method that takes account of this redundancy exists. Only $d+\binom{d+m-1}{m}$ multiplications are required. 
Implementing it showed it to be slower, so \ii\ does not use this idea.


\section{Log Signatures}\label{sec:logsigs}

Tensor space $T(\mathbb{R}^d)$, in which the signature of a $d$-dimensional path lives, has a notion of logarithm (\cite{FLA}, chapter 3), given by 
\begin{equation}\label{eq:log}
\log(1+T)=\sum_{n\ge1}\frac{(-1)^{n-1}T^n}{n}.
\end{equation}
Let $S$ be the set which consists of level $m$ of the signature of every path in $\mathbb{R}^d$.
$S$ is not the whole of the vector space $(\mathbb{R}^d)^{\otimes m}$, although it does \textit{span} $(\mathbb{R}^d)^{\otimes m}$
(see Lemma 8 in \cite{JD}). In fact, they form a lower-dimensional manifold. The 
logarithm operation 
maps this manifold continuously one-to-one to a linear subspace of $T(\mathbb{R}^d)$.
The representation of the logarithm of the signature in a basis of this subspace is called the \textbf{log signature}. 

The subspace in which the log signature of a path in $\mathbb{R}^d$ up to level $m$ lives is equivalent to the free $m$-nilpotent Lie algebra of type $d$, $\mathfrak{n}_{d,m}$. The log signature is like a compressed version of the signature up to the same level -- for every value in $\mathfrak{n}_{d,m}$, there is a path with that truncated log signature. The main source for the relevant mathematics is \cite{FLA} and an informal introduction is given in \cite{LOGSIG}.

$\mathfrak{n}_{d,m}$ is a finite dimensional real vector space, but there is no single obvious basis for it. In order to use the log signature as an efficient representation of a path, we need to choose a fixed basis.
There are two commonly used bases. They are both \emph{Hall bases}\cite{hall1950}. A Hall basis is made up of bracketed expressions, and it is determined by an ordering of all bracketed expressions.
\begin{itemize}
	\item The \emph{Lyndon basis}\cite{shirshov}, which is the default in \ii. Each basis element is labelled with a Lyndon word on $\{\alph1,\alph2,\dots,\alph{d}\}$, which is a sequence which comes earlier in lexicographic order than any of its \emph{rotations}. (For example, the rotations of $\alph{2432}$ are $\alph{2243}$, $\alph{3224}$ and $\alph{4322}$. $\alph{2243}$ and $\alph{1213}$ are Lyndon words but $\alph{31}$ and $\alph{3224}$ are not.) 
	\item The standard/canonical Hall basis, which we implement in such a way as to match CoRoPa\cite{coropa} exactly. 
	The ordering of equal-length expressions $[A,B]$ and $[C,D]$ is defined recursively: $[A,B]<[C,D]$ if either $A<C$ or ($A=C$ and $B<D$).
\end{itemize}

In these bases, each basis element is either a letter or a single bracketed expression, whose left and right are basis elements. We always pick an order on basis elements such that shorter bracketed expressions come before longer ones, and single letters, which are the first level, are in their natural order $\alph1<\alph2<\dots<\alph{d}$.

Much of the algebra calculations can be done once in the \verb|prepare| function. This is a major contribution of \ii\ and ensures for given $d$, $m$ and the choice of basis that the calculation is as efficient as possible. This is relevant in machine learning applications where typically many similar calculations are required.

\section{Log Signatures directly}\label{sec:c}

The log signature of a straight line displacement is just the displacement itself in level 1, and zero in every other level. The log signature of the concatenation of two paths is the Baker-Campbell-Hausdorff (BCH) product of the log signatures of the two paths. The direct method for calculating the log signature relies on being able to transform the log signature of a path given in terms of one of the bases above to the log signature of that path concatenated with a fixed line segment, achieved using the BCH product.

The BCH product is an infinite series in bracketed expressions in two indeterminates, which has can be formulated in different equivalent ways. The most straightforward ways express all brackets in the form of some Hall basis of the free Lie algebra of  $\mathbb{R}^2$. For example, using the Lyndon basis:
\begin{align*}
\mathrm{bch}(a,b)=a+b+\tfrac12[a,b]+\tfrac1{12}[a,[a,b]]+\tfrac1{12}[[a,b],b]+\tfrac1{24}[a,[[a,b],b]]+\dots.
\end{align*}

The coefficients in this expansion up to terms of depth twenty have been calculated and distributed by Fernando Casas and Ander Murua at \cite{bchinfo}, using their method described in \cite{bch}. We distribute their file as part of \ii, and read it when necessary. 

We can compute the Lie bracket of each pair of basis elements as a combination of other basis elements, and therefore, given two log signatures as combinations of basis elements (the second known to be just a displacement) we can find the expanded expression of their BCH product as a combination of basis elements. By doing this with indeterminates, the library develops an internal representation. 

As an example, in the case where the Lyndon basis is used, and we are concerned with two dimensions up to level two, a log signature looks like
\[a_0\alph1+a_1\alph2+a_2\alph{12}\]
The inductive step of the algorithm to accumulate log signatures by adding linear segments for $d=m=2$ is shown in Figure~\ref{fig:F22}. 

\begin{figure}[H]

\begin{lrbox}{\Lst}
\begin{lstlisting}[language=Python]
def F22(a, b):
    # Construct monomials of log signature a and displacement b
    t[0] = b[1] * a[0]
    t[1] = b[0] * a[1]
    # Extend log signature in-place
    a[2] += t[0] / 2    
    a[2] -= t[1] / 2
    a[0:2] += b[:]
\end{lstlisting}
\end{lrbox}
\begin{center}
\fbox{\usebox{\Lst}}
\end{center}
\caption{\label{fig:F22}Algorithm to accumulate a new displacement into a log signature in the Lyndon basis with $d=2$ and $m=2$.}
\end{figure}
If we go up to level 3, a log signature looks like
\[a_0\alph1+a_1\alph2+a_2\alph{12}+a_3\alph{112}+a_4\alph{122},\] with the final algorithm being as shown in Figure~\ref{fig:F23}.

\begin{figure}[H]
\begin{lrbox}{\Lst}
\begin{lstlisting}[language=Python]
def F23(a, b): # Log signature a and displacement b
    # Calculate monomials of a and b
    t[0]   += b[1] * a[0] # Order 2 monomials 
    t[1]   += b[1] * a[2]
    t[2]   += b[0] * a[1]
    t[3]   += b[0] * a[2]
    t[4]   += b[1] * t[0] # Order 3 monomials calculated from
    t[5]   += b[0] * t[0] #    t[i], i<=4
    t[6]   += b[1] * t[2]
    t[7]   += a[0] * t[0]
    t[8]   += a[1] * t[0]
    t[9]   += b[0] * t[2]
    t[10]  += a[0] * t[2]
    t[11]  += a[1] * t[2]
    # Extend log signature in-place
    a[2]   +=  t[0]/2 - t[2]/2
    a[3]   += -t[3]/2 - t[5]/12 + t[7]/12 + t[9]/12 - t[10]/12
    a[4]   +=  t[1]/2 + t[4]/12 - t[6]/12 - t[8]/12 + t[11]/12
    a[0:2] +=  b[:]
\end{lstlisting}
\end{lrbox}
\begin{center}
\fbox{\usebox{\Lst}}
\end{center}
\caption{\label{fig:F23}Algorithm to accumulate a new displacement into a log signature in the Lyndon basis with $d=2$ and $m=3$.}
\end{figure}
These functions have a lot of common structure. First a sequence of monomials in the input elements are constructed in the temporary array $t$. Higher order monomials are calculated inductively from other elements of $t$ to deduplicate the necessary multiplications.
Then some members of $a$ are incremented by some multiples of some of the temporary variables. Then the first $d$ elements of $a$ are incremented by all elements of $b$. Exactly which is given by the \verb|FunctionData| structure.
In general these functions are long and branching-free. The variable $a$ is modified in-place to produce the log signature of the extended path.

The basis (of the free Lie algebra on 2 symbols) used to express the BCH formula does not change the code we get, because the various equivalent bracketed expressions come to the same thing when they have been multiplied out. We use the Lyndon basis because it has slightly fewer terms, as \cite{bch} describes and partially explains. This choice is independent of the choice of basis (of the free Lie algebra on $d$ symbols) in which the log signature is expressed.
In general, we end up with fewer terms and a slightly faster calculation when the Lyndon basis is used for the log signature.

\section{Log Signatures from Signatures}\label{sec:s}
A simple method for calculating the log signature of a path is to calculate its signature first, and then convert to the log signature. The first step in doing the conversion is taking the logarithm itself in tensor space. This explicitly uses the formula (\ref{eq:log}) 
where $n$ only needs to go as high as the required level, and the power is in the concatenation product. This results in the log signature as an element of tensor space (which means it is as long as a signature), which is returned when \verb|logsig| is called with the \verb|"X"| (\emph{expanded}) method. The exact order of evaluation of formula (\ref{eq:log}) for best efficiency which we use is one which was suggested by Mike Giles\cite{Giles}.


To express this Lie element into a specified basis, we need to project it. We calculate a projection explicitly. There are known explicit forms for projections, for example the map given by the Dynkin-Specht-Wever lemma directly (\cite{DSWLemma}), which requires more operations. The \verb|prepare| function calculates a projection upfront.

Given the bracketed expression of a basis element with $m$ letters, we can easily find its expression in expanded space, by multiplying out the brackets. For example, $[[\alph1,\alph3],\alph3]$ is $\alph{133}-2\,\alph{313}+\alph{331}$. This gives us the full matrix $M_m$ to transform each level of the log signature to its expanded version. Each column of $M_m$ is labelled with a basis element, and each row is labelled with one of the $d^m$ words of length $d$. To compress level $m$ a given expanded log signature $x_m$ to its value $c_m$ in terms of a basis, we just need to solve a least squares problem $M_mc_m=x_m$. This problem is a very overdetermined system which is known to have an exact answer, up to rounding considerations. $M_m$ is tall and skinny. 

The words occurring in the terms of the expansion of such a bracketed expression are anagrams of the foliage of the expression. In that same example, for instance, $\alph{133}$, $\alph{313}$ and $\alph{331}$ are anagrams of $\alph{133}$. This leads to a lot of sparsity in the matrix $M_m$. Permuting the rows and columns to gather anagrams makes $M_m$ be a block diagonal matrix. We can save time doing the transformation by solving a separate linear system for each equivalence class of anagrams of words of length $m$.

For the standard Hall basis, this is exactly the procedure which we follow. In \verb|prepare|, we determine all the mapping matrices between anagram classes of the log signature and its expansion, and then we calculate all their Moore-Penrose pseudoinverses, so that solving the systems is just a matrix multiplication. 
The number of words in an anagram set containing $m$ letters where the frequency of the $i$th letter is $n_i$ is 
given by a multinomial coefficient $\frac{m!}{n_1!\dots n_d!}$. The number of Lie basis elements in an anagram set is given by the second Witt formula of Satz 3 of \cite{witt} as
\def\dummycommonfactor{\delta}
\def\lyn#1#2{\ensuremath{{\ell_{#1}}{\left(#2\right)}}}
\begin{equation}\label{eq:witt2}\lyn{m}{n_1,\dots,n_d}=\frac1m\sum_{\dummycommonfactor|n_i}\frac{\mu(\dummycommonfactor) (\frac{m}{\dummycommonfactor})!}{(\frac{n_1}{\dummycommonfactor})!\dots(\frac{n_d}{\dummycommonfactor})!},
\end{equation} where $\dummycommonfactor$ ranges over all common factors of the $n_i$ and $\mu$ is the M\"obius function. In the simple special case that the words have $m$ distinct letters, there are $m!$ words and $(m-1)!$ basis elements. In the Lyndon case, this formula makes sense because the Lyndon words in such a set of $m!$ words are just all that begin with the lowest letter. Typically the largest anagram sets are the ones with about the same number of each letter. For them, (\ref{eq:witt2}) is just $\frac1m$ times the number of words in the set because 1 is the only value of $\dummycommonfactor$. For example, looking at level 10 for a 3-dimensional path, the signature has 59049 elements and the log signature 5880, and there are 63 anagram classes.\footnote{The count is $63=\binom{10+3-1}{10}-3$ using the formula for unordered sampling with replacement and the fact that no basis element above level 1 has only one distinct letter in it.} The 12 most balanced anagram classes account for 3708 elements of the log signature, or $63.1\%$ of it. 

\begin{table}[H]
\begin{center}
\begin{tabular}{ccccc}
letter frequencies&number of classes&\multicolumn{1}{p{2cm}}{\centering signature\\elements\\in each}&\multicolumn{1}{p{2cm}}{\centering log signature\\elements\\in each}&\multicolumn{1}{p{2cm}}{\centering total log \\signature elements}\\
$\{4,3,3\}$&3&4200&420&1260\\
$\{4,4,2\}$&3&3150&312&936\\
$\{5,3,2\}$&6&2520&252&1512\\
$\{5,4,1\}$&6&1260&126&756\\
$\{6,2,2\}$&3&1260&124&372\\
\end{tabular}
\caption{\label{tab:anagramClasses}The sizes of the largest anagram classes for level 10 of $d=3$ in decreasing order of number of log signature elements. Many more such statistics have been tabulated in \cite{BLUMLEIN200419}.}
\end{center}
\end{table}
The big anagram classes account for most of the runtime when projecting to the log signature: multiplying a $420\times4200$ matrix by a 4200-vector takes 80\% more multiplications than multiplying a $312\times3150$ matrix by a 3150-vector and so on.

\subsection{Lyndon case}
If the Lyndon basis is required, then we have a more efficient implementation, which depends on a special property it has. On pages 89--91 of \cite{FLA}, the notation $P_a$ is introduced for the Lie polynomial corresponding to the Hall word $a$, i.e.~the polynomial you get by multiplying out the bracketed expression corresponding to the unique basis element whose foliage is $a$. This notation is used in the statement of the following.

\begin{theorem}[Theorem 5.1 of \cite{FLA}]
	The set of Lyndon words, ordered alphabetically, is a Hall set. The corresponding Hall basis has the following triangularity property: for each word $w=l_1\dots l_n$ written as a decreasing product of Lyndon words, the polynomial $P_w=P_{l_1}\dots P_{l_n}$ is equal to $w$ plus a $\mathbb{Z}$-linear combination of greater words.
\end{theorem}

The simplest case of the final statement, where $w$ is itself a single Lyndon word, gives the following useful fact. When the bracketed expression corresponding to a Lyndon word is expanded and terms are collected and ordered in alphabetical order of the word, the first term will be the Lyndon word itself, with coefficient 1. (For an example, consider the Lyndon word $\alph{133}$; its bracketed expression is $[[\alph1,\alph3],\alph3]$ and we saw earlier that this expands to $\alph{133}-2\,\alph{313}+\alph{331}$.) This means that the tall skinny matrix $M_m$ is lower triangular, as are its anagram blocks. If we take such a block and remove all the rows corresponding to words which are not Lyndon, we are left with the mapping from an anagram class in the compressed log signature to same Lyndon word elements of the expanded signature. It is a square lower triangular matrix with ones on the diagonal. We can now solve the system directly in many fewer operations, with just addition and multiplication, just looking at the Lyndon word elements of the expanded signature. \verb|prepare| determines the necessary indices and matrices, and \verb|logsig| does the solving.

For example, in level 4 on 3 dimensions, the following are the three basis elements which contain two $\alph1$s, a $\alph2$ and a $\alph3$:
\begin{align*}
[\alph1,[\alph1,[\alph2,\alph3]]]&=\alph{1123}-\alph{1132}-2\,\alph{1231}+2\,\alph{1321}+\alph{2311}-\alph{3211}\\
[\alph1,[[\alph1,\alph3],\alph2]]&=\alph{1132}-\alph{1213}+\alph{1231}-\alph{1312}-\alph{1321}+\alph{2131}-\alph{2311}+\alph{3121}\\
[[\alph1,\alph2],[\alph1,\alph3]]&=\alph{1213}-\alph{1231}-\alph{1312}+\alph{1321}-\alph{2113}+\alph{2131}+\alph{3112}-\alph{3121}
\end{align*}

The matrix corresponding to these looks as follows
\begin{align*}
\begin{blockarray}{cccc}
\rotatebox{-45}{$[\alph1,[\alph1,[\alph2,\alph3]]]$} & \rotatebox{-45}{$[\alph1,[[\alph1,\alph3],\alph2]]$} & \rotatebox{-45}{$[[\alph1,\alph2],[\alph1,\alph3]]$} \\
\begin{block}{(ccc)c}
1 & 0 & 0 & \alph{1123} \\
-1 & 1 & 0 & \alph{1132} \\
0 & -1 & 1 & \alph{1213} \\
-2 & 1 & -1 & \alph{1231} \\
0 & -1 & -1 & \alph{1312} \\
2 & -1 & 1 & \alph{1321} \\
0 & 0 & -1 & \alph{2113} \\
0 & 1 & 0 & \alph{2131} \\
1 & -1 & 0 & \alph{2311} \\
0 & 0 & 1 & \alph{3112} \\
0 & 1 & -1 & \alph{3121} \\
-1 & 0 & 1 & \alph{3211} \\
\end{block}
\end{blockarray}
\end{align*}
and when we restrict to Lyndon words (which in general are not the first rows) we get a matrix which has $m=4$ times fewer rows, and is a lower triangular square matrix with ones on the diagonal.

\nopagebreak 
\begin{minipage}{\textwidth}
\begin{align*}
\begin{blockarray}{cccc}
\rotatebox{-45}{$[\alph1,[\alph1,[\alph2,\alph3]]]$} & \rotatebox{-45}{$[\alph1,[[\alph1,\alph3],\alph2]]$} & \rotatebox{-45}{$[[\alph1,\alph2],[\alph1,\alph3]]$} \\
\begin{block}{(ccc)c}
1 & 0 & 0 & \alph{1123} \\
-1\tikzmark{mark1a} & 1 & 0 & \alph{1132} \\
0\tikzmark{mark3a} & -1\tikzmark{mark2a} & 1 & \alph{1213} \\
\end{block}
\end{blockarray}.
\end{align*}
If this matrix is called $M'$ we can solve the equation $M'c'=x'$ directly using
\begin{align*}
	c'_1=x'_1\qquad c'_2=x'_2-(-1\tikzmark{mark1b}\, c'_1)\qquad c'_3=x'_3-(0\tikzmark{mark3b}\,c'_1-1\tikzmark{mark2b}\, c'_2).
\begin{tikzpicture}[overlay,remember picture]
\draw[-{Latex},shorten >=8pt,shorten <=-2pt,out=350,in=130,distance=0.5cm,red, opacity=0.5] (mark1a.east) to (mark1b.north);
\draw[-{Latex},shorten >=6pt,shorten <=0pt,out=300,in=130,distance=0.5cm,red, opacity=0.5] (mark2a.east) -- (mark2b.north);
\draw[-{Latex},shorten >=8pt,shorten <=0pt,out=330,in=130,red, opacity=0.5] (mark3a.east) -- (mark3b.north);
\end{tikzpicture}
\end{align*}
\end{minipage}

\section{Implementation}\label{sec:impl}
\ii\ is a Python package which is built on \numpy\cite{numpy}, which is ubiquitous for dealing with numerical data in Python. 
The Python ecosystem is very commonly used for deep learning.
It 
is 
implemented as a \CC\ extension. 

There is a single \verb|.cpp| file which defines the whole interface with Python. The mathematical functionality resides in header files. This \emph{unity build} structure reduces the time to build the whole library, which matters to users, at the cost of incremental build time.

\subsection{Signatures}
We store signatures during the calculation with each level in a contiguous block of memory, which means that accesses are efficient. 
We start with the signature of the first displacement and step-by-step concatenate on the signature of each succeeding displacement. The concatenation is done in place, but in simple cases this doesn't seem to make a difference in performance.

We also wrote an implementation of the signature calculation using a template metaprogramming style, where the dimension and level are template parameters, there is no heap memory allocation and all loops are constant length. We compared the methods and learnt that the performance is the same. Because we want to allow arbitrary calculations easily for the user, it is convenient not to code in this way inside iisignature.
\subsection{Preparing the direct calculation of log signatures}
The internal representation of the calculation required to convert the log signature of a path into the log signature of that path with a line segment concatenated on the end is stored in an instance of the \verb|FunctionData| class. The calculation depends on the following concepts. The class \verb|Input| represents an indeterminate, and a \verb|Coefficient| is a polynomial in \verb|Input|s 
Elements of the basis of the free Lie algebra we are using are represented by instances of the \verb|BasisElt| class. These are created once for a whole calculation, and they all live together in memory controlled by a \verb|BasisPool| which also remembers their order and those of their Lie products (Lie brackets) which happen to be \verb|BasisElt|s. Elements of the free Lie algebra, Lie polynomials, 
 are represented by the class \verb|Polynomial|. 
In a similar way as \cite{coropa}, we store the data of a \verb|Polynomial| with each level separately, this speeds up the multiplication of two of them very much, because it becomes trivial to avoid trying to multiply terms whose combined level will exceed the level we are truncating at. The procedure calculates the BCH product of an arbitrary polynomial (one with a separate indeterminate for each basis element, representing an arbitrary log signature) and an arbitrary level-one polynomial (one which is just a separate indeterminate for each letter, representing the log signature of a single displacement) to produce a \verb|Polynomial| which is exactly what the \verb|FunctionData| needs to calculate.

\begin{table}[H]
\begin{center}
\renewcommand{\arraystretch}{1.7}
\begin{tabular}{ llll}
\hline
type    &    algebraic structure &   definition &instance represents\\
\hline
\verb|double|&field (roughly)&\CC\ builtin&constant floating point real\\
\verb|Input|&set&indeterminate&numeric input\strut\\
\verb|Coefficient|&semigroup ring& \parbox[t]{3.7cm}{\raggedright\strut polynomial from $\texttt{double}[\{\texttt{Input}\text{s}\}]$
}&\parbox[t]{3.7cm}{\raggedright\strut formula in terms of the inputs}\\
\verb|BasisElt|&set&(fixed but complicated)&\parbox[t]{3.7cm}{\raggedright\strut element of the given basis of the FLA}\\[3pt]
\verb|Polynomial|& 
\parbox[t]{3cm}{
\raggedright\strut free vector space augmented with Lie bracket\strut}
& \parbox[t]{3.7cm}{\raggedright function from \texttt{BasisElt} to \texttt{Coefficient}
\strut}
& element of FLA\\
\hline
\end{tabular}
\caption{\label{tab:bchobjects}Summary of the main object types for the Free Lie algebra (FLA) calculations}
\end{center}
\end{table}

In the \verb|"O"| (\emph{object}) mode, the \verb|prepare| function goes as far as computing this \verb|FunctionData| object, and the \verb|logsig| function follows its instructions for dealing with each displacement, using the function \verb|slowExplicitFunction|.



\subsection{On-the-fly machine code generation for the direct calculation}

Code specifically compiled for the particular function is more efficient than following instructions given by the \verb|FunctionData| object. Before this library, we wrote some code (described in \cite{LOGSIG} and demonstrated at \url{https://github.com/bottler/LogSignatureDemo}) to generate \CC\ code which can be compiled to give efficient versions of this function. We learnt that while this method is very efficient, it is impractical for many realistic $d$ and $m$ because the function can easily get so large that compilers take unreasonably long times to compile it. Attempting to split them up only helps a small amount. Manual machine code generation avoids this delay. \ii\ therefore provides the \verb|"C"| method under which it compiles the \verb|FunctionData| itself on-the-fly to machine code internally in a buffer in memory during \verb|prepare|, and all \verb|logsig| need do is run the compiled code for each displacement. This is implemented for x86 and x86-64 architectures, for Windows, Linux and Mac.

The logic for the compilation is in \verb|makeCompiledFunction.hpp|. 
The \verb|Mem| object represents a buffer for storing machine code, which is allocated in such a way that execution is enabled.
The \verb|FunctionRunner| object is constructed from a \verb|FunctionData| and allocates a \verb|Mem| and compiles the function into it, providing a \verb|go()| function to run the compiled code. The actual compilation is done by the \verb|Maker| class. The comments in that class explain what it is doing in terms of 
\verb|x86| and \verb|x86-64|
 machine code instructions. In the \verb|x86| case, we rely on SSE2 instructions for floating point arithmetic 
which enables the logic to be roughly the same as in the 64-bit case. 


\subsection{Projection from expanded log signature to a basis}
The \verb|makeMappingMatrix| function calculates the full matrix (in sparse form) to project from tensor space to the desired basis. The identification of anagram classes is performed in \verb|analyseMappingMatrixLevel|. The function \verb|makeSparseLogSigMatrices| identifies all the data needed to do the projection from this information. When, as often happens, 
a basis element's anagram class is a singleton, we can just read off its value from the expanded log signature without solving a system. 
In the standard Hall basis case, we need to calculate the Moore-Penrose pseudoinverses of the identified matrices, and we do this using \numpy\ at the interface level. All the data to do this is stored in the 
object which \verb|prepare| generates, and is available to be simply used by \verb|logsig|.

\section{Indicative timings}\label{sec:time}
Using 64bit Python 3.5 on Ubuntu 16.04LTS with an AMD FX-8320, timings were taken for calculating 100 signatures of randomly generated paths with 100 timesteps in various different ways. We compare here a native python signature implementation using \numpy, the calculation with \ii\ version 0.20,  and the calculation in the package \verb|esig.tosig| of CoRoPa\cite{coropa}, version 0.6.5. Both \ii\ and \verb|esig| use 64-bit floating point internally, but \ii\ is taking and returning 32-bit floating point values in this example, whereas \verb|esig| uses 64 bit throughout. 
It should be noted that \verb|esig| was not specifically written to make this type of calculation fast, but for other types of flexibility (e.g.~the sparse signature case). The dramatic difference shows the advantage of having code written specifically for the dense case. 
\newcolumntype{H}{>{\iffalse}c<{\fi}@{}} 

\begin{figure}[H]
\begin{center}
	\includegraphics[width=4.387in]{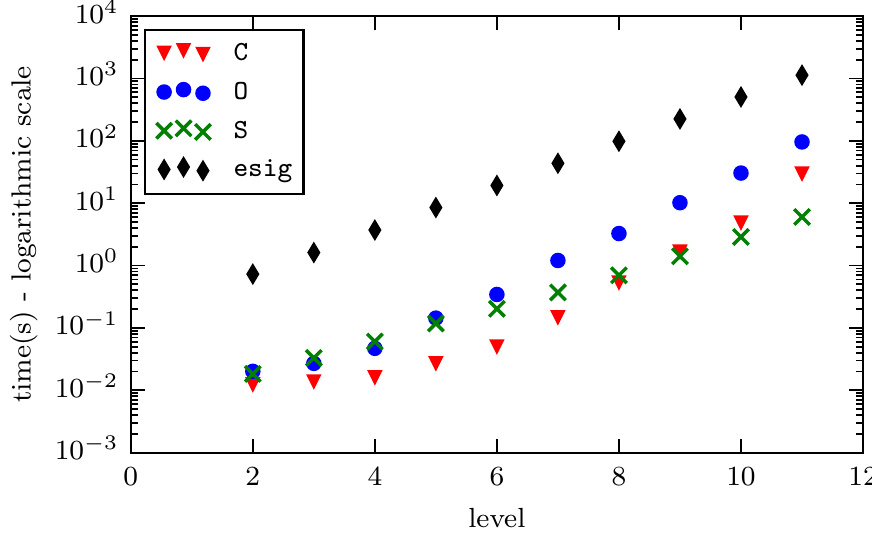}
	\caption{\label{fig:logsigtiming2d}Various log signature calculation timings in seconds, for 1000 random 2-dimensional paths of 100 steps each for various levels. For \ii, only the Lyndon basis is shown.}
\end{center}
\end{figure}

\begin{figure}[H]
\begin{center}
	\includegraphics[width=4.387in]{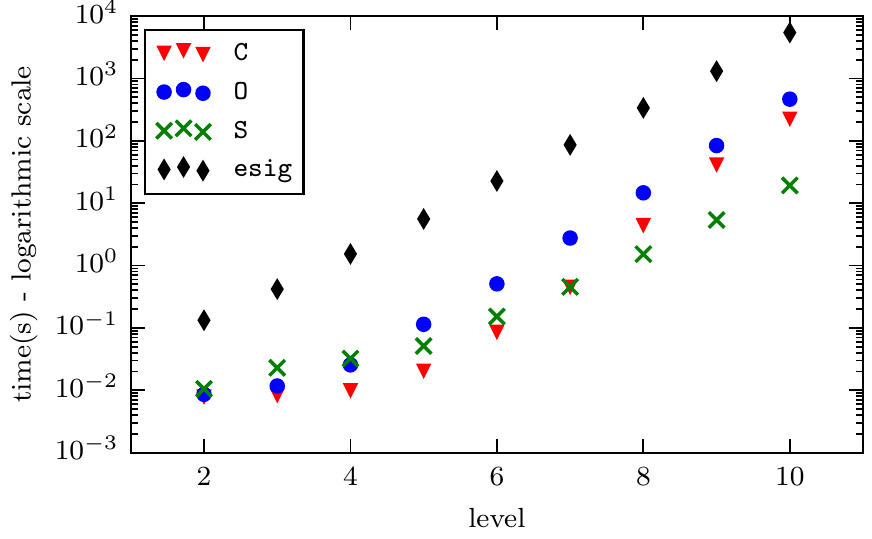}
	\caption{\label{fig:logsigtiming3d}Various log signature calculation timings in seconds, for 1000 random 3-dimensional paths of 10 steps each for various levels. For \ii, only the Lyndon basis is shown.}
\end{center}
\end{figure}

\begin{table}[H]
\begin{center}
\begin{tabular}{ l rrrrrr}
	\hline
	\hfill($d$,$m$)             &    (2,6) &   (2,10) & (3,10)   &   (5,5) & (10,4)   \\
	\hline
	Python native         &   10.56 &   78.66 & 2458.09 &  55.95 & 118.83  \\
	\verb|iisignature.sig|   &   \bftab{0.02} &    \bftab{0.27} & \bftab{10.78}    &   \bftab{0.29} & \bftab{0.61}    \\
	\verb|esig.tosig.stream2sig|    &    1.98 &    55.51 & 3114.36  &   56.36 & 175.38   \\
	\hline
\end{tabular}
\caption{\label{tab:sigtiming}Various signature calculation timings in seconds, for 100 random paths of 100 steps each in the given combinations of level and dimension}
\end{center}
\end{table}
Calculating the log signature using the compiled method is quicker than calculating the signature for small depths, but for larger depths the signature becomes significantly faster. 
As the depth increases, the projection method therefore becomes the best method to obtain the log signature. For two dimensions, performance is shown in Table~\ref{tab:logsigtiming2d} and plotted in Figure~\ref{fig:logsigtiming2d}, and for three dimensions performance is shown in Table~\ref{tab:logsigtiming3d} and plotted in Figure~\ref{fig:logsigtiming3d}. 
\begin{table}[H]
\begin{center}
\begin{tabular}{rlHrrrrrrrrr}
	\hline
&level:&2&3&4&5&6&7&8&9&10&11\\
	\hline
	\verb|C|&Lyndon  &     \bftab{0.01} &     \bftab{0.01} &     \bftab{0.02} &     \bftab{0.03} &     \bftab{0.05} &     \bftab{0.14} &     \bftab{0.52} &     1.64 &      4.90&29.21 \\
	\verb|C|&\rlap{standard} &     \bftab{0.01} &     \bftab{0.01} &     \bftab{0.02} &     \bftab{0.03} &     0.05 &     0.15 &     0.55 &     1.74 &      5.31&32.34  \\
	\verb|O|&Lyndon  &     0.02 &     0.03 &     0.05 &     0.14 &     0.34 &     1.21 &     3.26 &     10.18 &     30.46&95.98 \\
	\verb|O|&\rlap{standard} &     0.02 &     0.03 &     0.05 &     0.15 &     0.36 &     1.25 &     3.38 &    10.80 &     32.89&{107.35}\\
	\verb|S|&Lyndon  &     0.02 &     0.03 &     0.06 &     0.12 &     0.20  &     0.37 &     0.70 &     \bftab 1.41 &      \bftab 2.87&{\bftab 6.01} \\
	\verb|S|&\rlap{standard} &     0.02 &     0.04 &     0.06 &     0.11 &     0.21 &     0.37 &     0.70 &     1.43 &       2.92&6.15 \\
	\verb|esig|&\rlap{standard}      &     0.73 &     1.61 &     3.71 &     8.52 &    19.26 &    43.66 &    98.47 &   224.83 &    506.25&\llap{1132.24}\\
	\hline
\end{tabular}
	\caption{\label{tab:logsigtiming2d}Various log signature calculation timings in seconds, for 1000 random 2-dimensional paths of 100 steps each for the given levels}
\end{center}
\end{table}
\begin{table}[H]
\begin{center}
\begin{tabular}{rlrrrrrrrrrr}
	\hline
&level:&2&3&4&5&6&7&8&9&10\\
	\hline
\verb|C|    & Lyndon   & \bftab 0.01 & \bftab 0.01 & \bftab 0.01 & \bftab 0.02 & \bftab 0.08 & \bftab 0.45 & 4.35        & 40.80       & 221.41       \\
 \verb|C|    & standard & 0.01        & 0.01        & 0.01        & 0.02        & 0.09        & 0.50        & 5.36        & 47.11       & 255.75       \\
 \verb|O|    & Lyndon   & 0.01        & 0.01        & 0.03        & 0.11        & 0.51        & 2.76        & 14.67       & 84.04       & 466.12       \\
 \verb|O|    & standard & 0.01        & 0.01        & 0.02        & 0.12        & 0.53        & 2.98        & 16.50       & 101.06      & 605.05       \\
 \verb|S|    & Lyndon   & 0.01        & 0.02        & 0.03        & 0.05        & 0.15        & 0.46        & \bftab 1.52 & \bftab 5.38 & \bftab 19.25 \\
 \verb|S|    & standard & 0.01        & 0.02        & 0.03        & 0.05        & 0.15        & 0.51        & 1.92        & 8.36        & 41.06        \\
 \verb|esig| & standard & 0.13        & 0.42        & 1.54        & 5.59        & 22.69       & 86.24       & 338.08      & 1310.86     & 5451.14      \\
	\hline
\end{tabular}
	\caption{\label{tab:logsigtiming3d}Various log signature calculation timings in seconds, for 1000 random 3-dimensional paths of 10 steps each for the given levels}
\end{center}
\end{table}

\begin{figure}[H]
\begin{center}
	\includegraphics[width=4.387in]{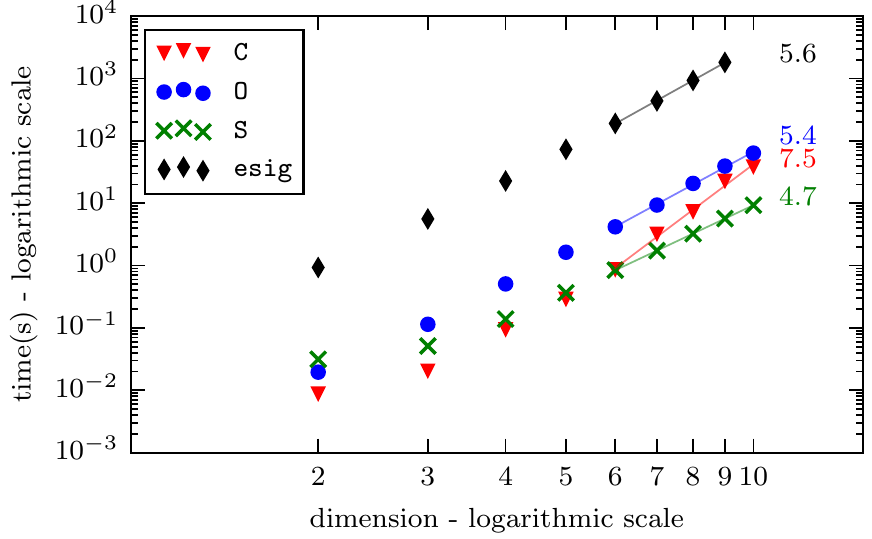}
	\caption{\label{fig:logsigtimingm5}Various level-5 log signature calculation timings in seconds, for 1000 random paths of 10 steps each of various dimensions. For \ii, only the Lyndon basis is shown. The graphs look to have roughly reached a straight line for $d\ge6$. The least squares line of each is shown, with its gradient which indicates the approximate degree of a polynomial relationship.}
\end{center}
\end{figure}

We expect the time taken to increase polynomially in dimension, so we plot the time taken for various methods as $d$ increases on a log-log plot in Figure~\ref{fig:logsigtimingm5} and show the timings in Table~\ref{tab:logsigtimingm5}. Ultimately the calculation time would be expected to be quintic in $d$. For $d$ in the high single figures, we observe much higher growth in the runtime of the compiled code.

\begin{table}[H]
\begin{center}
\begin{tabular}{rlrrrrrrrrr}
\hline
&\rlap{dimension:}&2&3&4&5&6&7&8&9&10\\
\hline
 \verb|C|    & Lyndon   & \bftab 0.01 & \bftab 0.02 & \bftab 0.09 & \bftab 0.29 & 0.85        & 3.17        & 7.24        & 22.33       & 37.83       \\
 \verb|C|    & standard & 0.01        & 0.02        & 0.10        & 0.30        & 1.03        & 3.35        & 7.58        & 21.64       & 38.38       \\
 \verb|O|    & Lyndon   & 0.02        & 0.11        & 0.51        & 1.63        & 4.18        & 9.34        & 20.76       & 39.39       & 63.67       \\
 \verb|O|    & standard & 0.02        & 0.12        & 0.52        & 1.69        & 4.38        & 9.84        & 19.72       & 38.44       & 64.92       \\
 \verb|S|    & Lyndon   & 0.03        & 0.05        & 0.14        & 0.37        & \bftab 0.84 & \bftab 1.73 & \bftab 3.24 & \bftab 5.66 & \bftab 9.30 \\
 \verb|S|    & standard & 0.03        & 0.05        & 0.15        & 0.41        & 0.96        & 2.02        & 4.56        & 6.81        & 11.60       \\
 \verb|esig| & standard & 0.93        & 5.59        & 22.75       & 73.17       & 190.47      & 437.35      & 931.58      & 1814.74     &             \\
\hline
\end{tabular}
	\caption{\label{tab:logsigtimingm5}Various level-5 log signature calculation timings in seconds, for 1000 random paths of 10 steps each of various dimensions.}
\end{center}
\end{table}

\begin{figure}[H]
\begin{center}
	\includegraphics[width=4.387in]{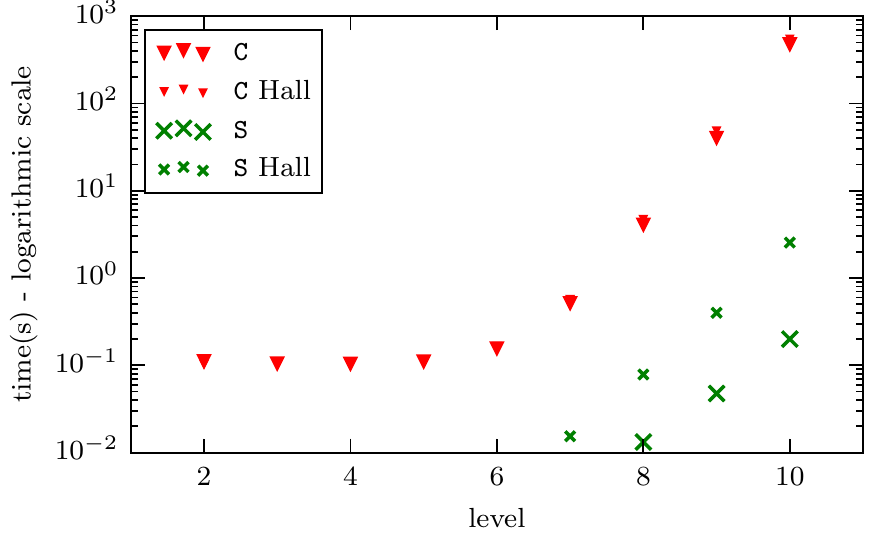}
	\caption{\label{fig:preptiming3d}Timings for a single preparation of the 3-dimensional log signature calculation for various levels. Smaller marks are used for the standard Hall basis, regular marks for the Lyndon basis. Values for \texttt{O} and \texttt{C} are very similar, so the former are omitted. Very small values are also omitted.}
\end{center}
\end{figure}

\begin{figure}[H]
\begin{center}
	\includegraphics[width=4.387in]{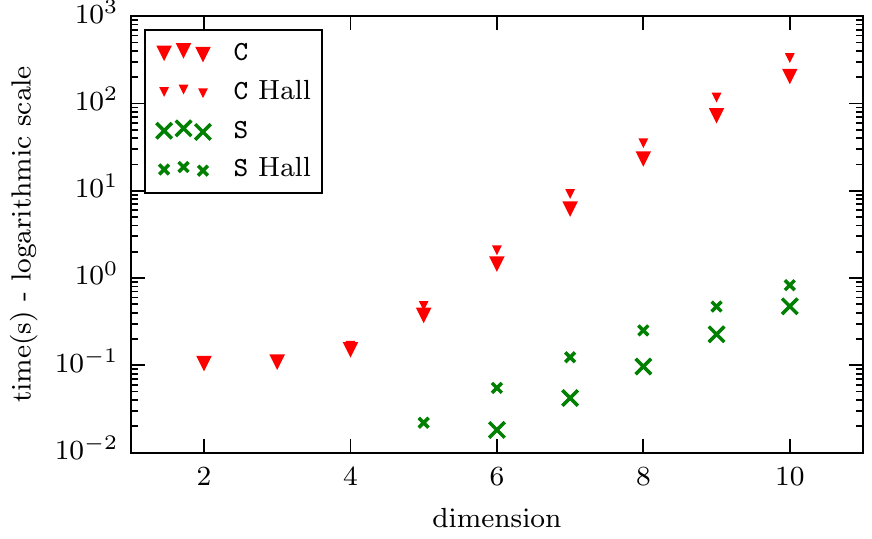}
	\caption{\label{fig:preptimingm5}Timings for a single preparation of the level-5 log signature calculation for various dimensions. Only the Lyndon basis is shown. Smaller marks are used for the standard Hall basis, regular marks for the Lyndon basis. Values for \texttt{O} and \texttt{C} are very similar, so the former are omitted. Very small values are also omitted.}
\end{center}
\end{figure}

The preparation in \ii\ is slow for the \verb|C| method when $d$ or $m$ is large. Timings for a single call are illustrated for various levels with $d=3$ in Figure~\ref{fig:preptiming3d} and for various dimensions with $m=5$ in Figure~\ref{fig:preptimingm5}. There is an advantage in using the Lyndon basis.

While there is always more that can be done to speed software performance, \ii\ provides a significant speedup over other options easily available to those using python for machine learning and doing lots of signature calculations.
\section{Indicative memory usage}\label{sec:mem}
We used the Massif tool \cite{massif} from Valgrind to profile memory usage calculating a single log signature, in particular collecting the peak memory usage. The peak memory usage is interesting because running out of memory is usually what makes certain calculations impossible. There is a background memory cost independent of the algorithm, which includes space to store the BCH coefficients. In order just to measure the algorithm, we ran these calculations from within \CC\ without using Python, and we subtract the memory usage calculating the same signature at level 1 from the observed memory usage. 

\begin{figure}[H]
	\begin{center}
		\includegraphics[width=4.387in]{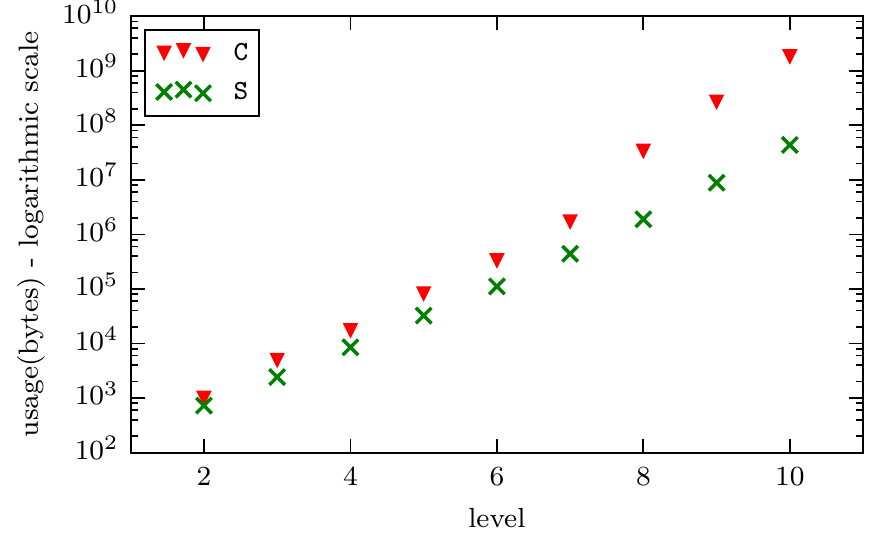}
		\caption{\label{fig:logsigmem3d}Memory usages in bytes for \texttt{C} and \texttt{S} calculations of a three-dimensional path of 10 steps for various levels. Only the Lyndon basis is shown.}
	\end{center}
\end{figure}

The values are very consistent across repeated runs. Values for three-dimensional paths are shown in Figure~\ref{fig:logsigmem3d}. We observe that memory usage is another reason why the projection method becomes a better choice for higher levels. 

\section{Other functionality}\label{sec:other}
We have described the core functionality of \ii, the calculation of signatures and log signatures. There are also simple methods for concatenating a segment onto a signature, and transforming a given signature according to the path being scaled by an enlargement in each of the coordinate directions. Counterparts for backpropagating derivatives through the signature calculation and through these are provided, allowing them to occur in the middle of a neural network.
The library also implements functions for calculating the linear combinations of signature elements of 2d paths which are invariant under rotation, which were derived in \cite{JD} and have been applied to handwriting recognition.

\section{Conclusions and future work}

We have presented and analysed efficient methods for computing signatures and log signatures. 
We have implemented what we considered to be the most promising algorithms.

We have focused on small-dimensional data in our design, because this encompasses many applications where the signature has been used in machine-learning. For example handwriting recognition and EEG data.
Calculating the log signature in cases where $d\gg m$ has not been a priority. For example $d>50$ and $m\le4$. Some data is naturally a high-dimensional sequence, possibly discrete (making movements of a fixed length in a single dimension at a time), like some representations of music and text, so this is a possible use case. In these cases there is lots of repetition in the calculations. There are potential changes to the code which would make \verb|prepare| use significantly less memory (and therefore be usable for larger $d$) in this regime, at the cost of a little more calculation time in \verb|logsig|.

We have shown that machine code generation directly from the algebra is useful in this domain.
The calculation is data-access heavy and the order of operations has a big effect, because of memory latency and data dependencies, and there is scope to improve it. We find that adding extra operations to the code without changing the data access does not slow it down, suggesting that it is the effect of data-cache misses which is the main bottleneck. There are subsets of the calculation which are repeated on different parts of the data. Operating in parallel with vector instructions might speed things up.
There are avenues for working on these possibilities, for example using the LLVM system, which brings the advantages of a modern compiler to the code generation. 
\section{Acknowledgements}
We thank Terry Lyons and Joscha Diehl for many helpful suggestions.

\printbibliography
\end{document}